\documentclass[showpacs,aps]{revtex4}
\usepackage{epsfig,amsmath}

\begin{document}

\title{Chaotic Motion Around Stellar Objects with Octupolar Deformation:\\
Newtonian and Post Newtonian Approaches}

\author{Javier Ramos-Caro}\email[e-mail: ]{javiramos1976@gmail.com}

\author{Framsol L\'opez-Suspes}\email[e-mail: ]{framsol@gmail.com}

\author{Guillermo A. Gonz\'alez}\email[e-mail: ]{gonzalez@gag-girg-uis-net}

\affiliation{Grupo de Investigaci\'on en Relatividad y Gravitaci\'on, Escuela de
F\'{\i}sica, Universidad Industrial de Santander, A.A. 678, Bucaramanga,
Colombia}


\begin{abstract}

Regular and chaotic test particle motion in gravitational fields due to stellar
bodies with quadrupolar and octupolar deformation are studied using Poincar\'e
surfaces of section. In first instance, we analyze the purely Newtonian case and
we find that the octupolar term induces a distortion in the KAM curves
corresponding to regular trajectories as well as an increase in chaoticity, even
in the case corresponding to oblate deformation. Then we examine the effect of
the first general relativistic corrections, provided by the post Newtonian
approach. For typical values of the post Newtonian multipoles we find that the
phase-space structure practically remains the same as in the classical case,
whereas that for certain larger values of these multipoles the chaoticity
vanishes. This important fact provides an interesting example of a situation
where a non-integrable dynamical system becomes integrable through the
introduction of a large perturbation.

\end{abstract}

\maketitle

\section{Introduction}\label{intro}

As is suggested by a wide variety of observational evidences, many astrophysical
objects can be modeled as axially symmetric bodies with prolate or oblate
deformation. For example, it is known that the Earth has non vanishing
quadrupolar and octupolar moments, as a consequence of its oblate
shape\cite{puccaco}. Also, many galaxies with a large disc component can be
assumed as axisymmetric oblate bodies with a large quadrupolar moment and, in
some cases, with a comparable octupolar deformation due to the remaining
components (for example, the halo). Likewise, there are galaxy clusters with a
cigarlike shape\cite{Cooray} and many dwarf galaxies that can be considered as
nearly axisymmetric prolate deformed objects\cite{ryden}. Although in the above
cases the quadrupolar moment it is considered to be the major deviation from the
spherical symmetry, there are situations (for example, some metallic clusters)
where the octupolar deformation play a significant
role\cite{hamamoto,frauendorf}.

The motion of test particles around such stellar objects is a problem of wide
physical interest. The case of attraction centers described by monopolar plus
quadrupolar terms has been extensively studied by Guer\'on and Letelier, from a
classical and relativistic standpoint, showing that the inclusion of external
multipolar moments can induce chaos\cite{letelier,letelier2,letelier3}. In
Newtonian gravity, as well as in general relativity, chaos can be found when the
source has prolate deformation and the chaoticity grows by increasing the
quadrupolar moment. On the other hand, it seems to be that the case
corresponding to oblate deformation does not lead to chaotic motion, indeed for
a very large quadrupolar deformation. However, inclusion of octupolar
deformation can also induce chaos, as it was shown by Heiss, Nasmitdinov and
Radu\cite{heiss} and Li\cite{li} in the case of an harmonic oscillator.

In the present paper we investigate the motion of test particles in axially
symmetric gravitational potentials that are the sum of a monopole, a quadrupole
and an octupolar term (only external multipole moments will be considered). At
the first instance, in Sec. \ref{newton}, we shall perform the analysis in the
context of Newtonian gravity by examining how the structure of the Poincar\'e
surfaces of section is determined by the octupolar moment. We find that, even in
the case of oblate deformation, modest values of the octupolar moment induce
chaotic motion. Then, in order to compare how these results are modified by
considering the first general relativistic corrections, we shall introduce in
Sec. \ref{1PN} the post Newtonian (1PN) approach, where the gravity is described
by the classical Newtonian potential and another additional field, which is
written in terms of the 1PN multipolar moments. Contrary to the usual statement
that integrable problems in Newtonian theory become non integrable in general
relativity, we find that, for certain values of the 1PN multipolar moments, the
test particle motion is regular in situations where the Newtonian theory
predicts chaos.

\section{Transition Ring-spindle Torus and Chaos Induced by Octupolar Deformation}\label{newton}

Consider a test particle moving in the axially symmetric gravitational field
generated by a stellar body with quadrupolar and octupolar deformation. In
cylindrical coordinates $(R,z,\varphi)$, its potential has the form
\begin{equation}
\Phi=-\frac{\alpha}{\sqrt{R^{2}+z^{2}}}-\frac{\beta
(2z^{2}-R^{2})}{2(R^{2}+z^{2})^{5/2}}-\frac{\gamma
(2z^{3}-3zR^{2})}{2(R^{2}+z^{2})^{7/2}}\label{multipolarNewton}.
\end{equation}
Since we are interested on to describe the exterior test particle motion, we
consider only external multipolar moments. In the above equation $\alpha$ is the
monopole term that equals to $Gm$, where $m$ is the total mass of the source and
$G$ is the gravitational constant. The quadrupolar term, denoted by $\beta$,
usually represents the major deviation from spherical symmetry. In particular,
if $\beta >0$ the source has prolate deformation and if $\beta <0$ we have the
case corresponding to oblate deformation. The octupolar moment $\gamma$
describes the asymmetry of the source with respect to the equatorial plane, i.e.
its ``shape-of-pear'' deformation. Both $\beta$ and $\gamma$ are related to the
source's density $\rho(R,z)$ through the equations \cite{bt}
\begin{eqnarray}
  \beta & = & 2\pi\int_{0}^{\infty}r'^{4} dr'\int_{0}^{\pi}d\theta '\sin\theta '
  P_{2}(\cos \theta ')\rho(r',\theta '),\label{cuadrupoloNew}\\
   &  &  \nonumber\\
  \gamma & = & 2\pi\int_{0}^{\infty} r'^{5}dr'\int_{0}^{\pi}d\theta '\sin\theta
  ' P_{3}(\cos \theta ')\rho(r',\theta '),\label{octupoloNew}
\end{eqnarray}
where we have used spherical coordinates $r=\sqrt{R^{2}+z^{2}}$,
$\cos\theta=z/\sqrt{R^{2}+z^{2}}$ and $P_{l}$ denotes the Legendre polinomial of
order $l$.

The motion of a test particle in a gravitational field described by
(\ref{multipolarNewton}), obeys the relations
\begin{eqnarray}
\dot{R} &=& {V}_{R}, \label{em1} \\
\dot{z} &=& {V}_{z}, \\
\dot{V}_{R} &=& - \frac{\partial {\Phi}_{ef}}{\partial R}, \\
\dot{V}_{z} &=& -
\frac{\partial {\Phi}_{ef}}{\partial z}, \label{em2}
\end{eqnarray}
where ${\Phi}_{ef}$ is the effective potential, given by
\begin{equation}
{\Phi}_{ef} = \Phi + \frac{\ell^{2}}{2R^{2}}.\label{potefNewton}
\end{equation}
Here, $\ell=R^{2}\dot{\varphi}$ is the axial specific angular momentum that is
conserved as a consequence of the axial symmetry. The second integral of motion
is the total specific energy, given by
\begin{equation}
E =({V}_{R}^{2}+{V}_{z}^{2})/2 + {\Phi}_{ef}.\label{EnrgyNewton}
\end{equation}
According to eqs. (\ref{em1})-(\ref{EnrgyNewton}), the motion is restricted to a
three dimensional phase space $(R,z,V_{R})$. This fact enable us to introduce
the Poincar\'e surfaces of section method, in order to investigate the
trajectories of test particles.

\begin{figure}
  \epsfig{width=9cm,file=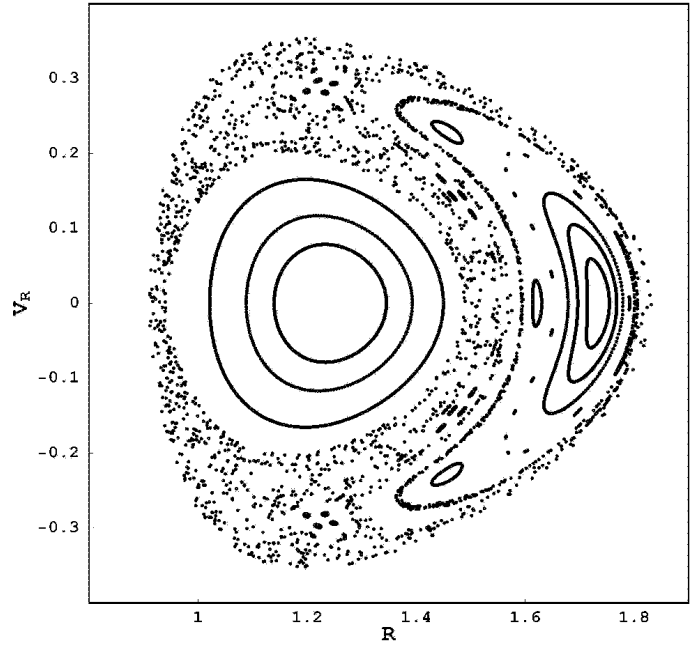}\\
  \caption{Surface of section for some orbits with $\ell=0.9$,
  $E=-0.4$, in a potential characterized by $\alpha=1$, $\beta=0.3$
  and $\gamma=0$.}\label{quadrupolar1}
\end{figure}

\begin{figure}
  \epsfig{width=9cm,file=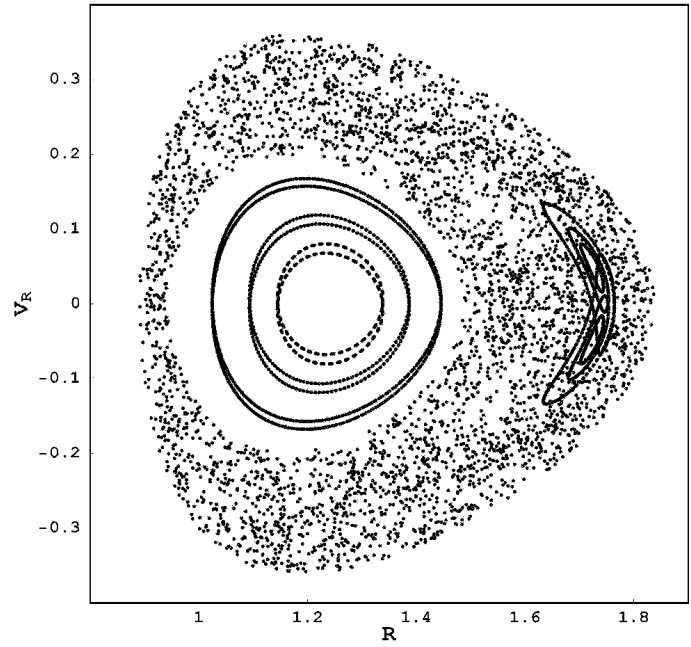}\\
  \caption{Surface of section for the same initial conditions of the previous
  figure. We maintain the values $\ell=0.9$,
  $E=-0.4$, $\alpha=1$ and $\beta=0.3$, but now $\gamma=0.02$.}\label{octopolar1a}
\end{figure}

\begin{figure}
  \epsfig{width=9cm,file=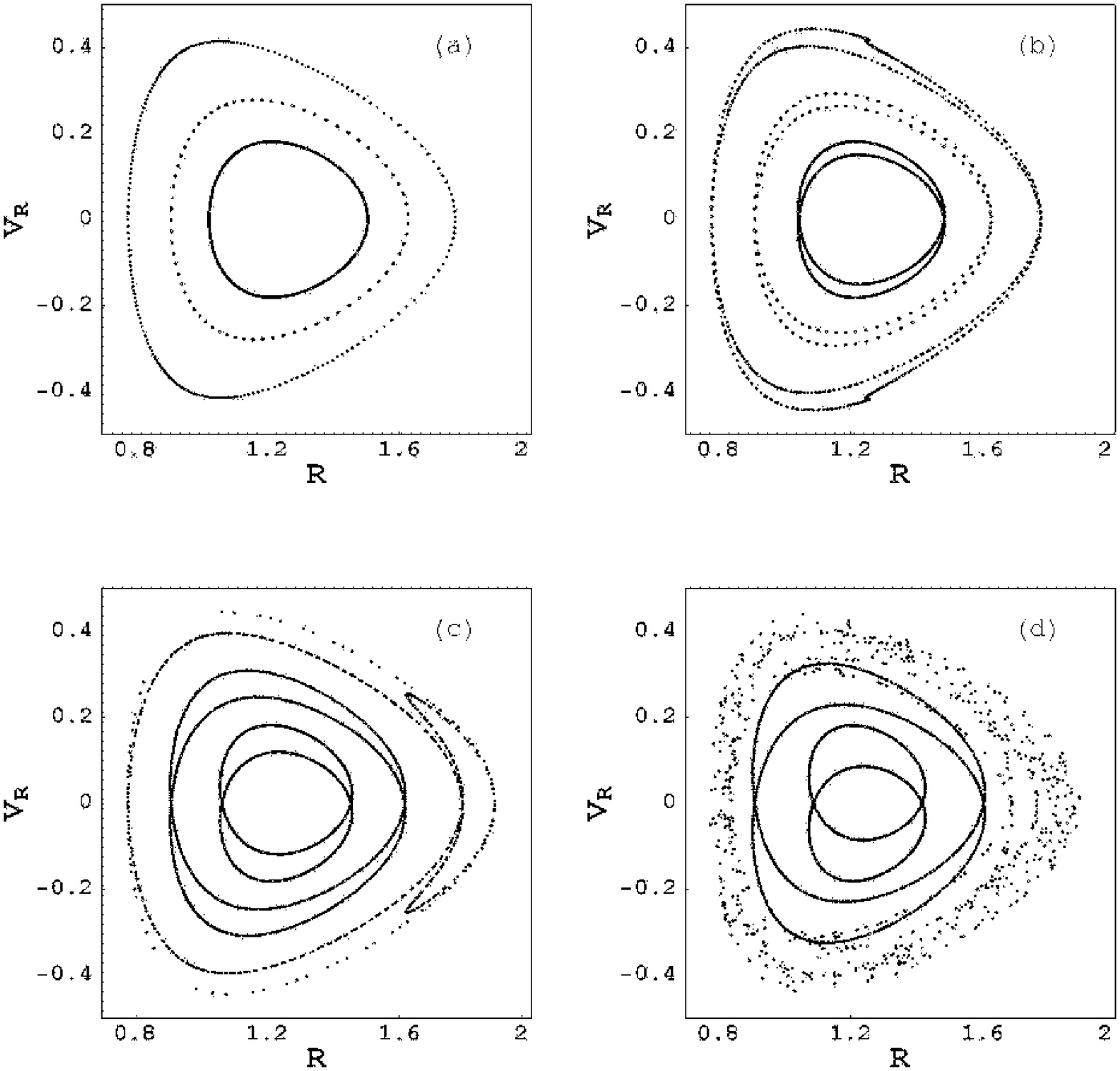}\\
  \caption{Surfaces of section for $\ell=0.9$, $E=-0.4$, $\alpha=1$, $\beta=0.2$ and
  (a) $\gamma=0$; (b) $\gamma=0.02$; (c) $\gamma=0.04$; (d)
  $\gamma=0.06$. In each case, they are generated by three orbits with initial
  conditions
(i) $z=0$, $R=0.91$, $V_{R}=0$; (ii) $z=0$, $R=0.78$, $V_{R}=0$ and
(iii) $z=0$, $R=1.16$, $V_{R}=0.18$.}\label{octupolar2}
\end{figure}

\begin{figure}
  \epsfig{width=9cm,file=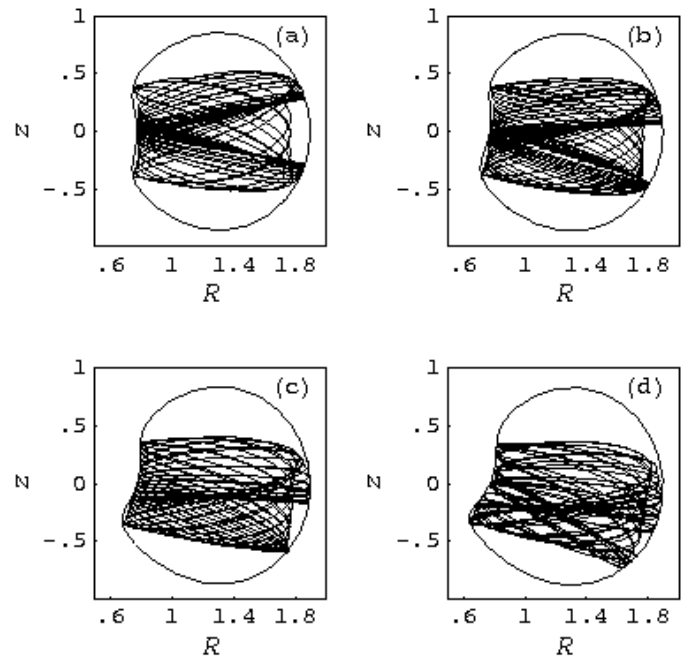}\\
  \caption{An orbit in the meridional plane with initial conditions
  $z=0$, $R=0.78$, $V_{R}=0$ and the same parameters considered above.
  Again we have the cases (a) $\gamma=0$; (b) $\gamma=0.02$; (c) $\gamma=0.04$; (d) $\gamma=0.06$}
  \label{meridionaloctupolar2}
\end{figure}

In Fig. \ref{quadrupolar1} we plot a typical $z=0$ surface of section
corresponding to test particle motion in the presence of a gravitational field
due to a prolate deformed source whom octupolar moment vanishes (from here on,
we choose $\alpha=1$, without loss of generality). We note a central and lateral
regular regions composed by ring torus (ring KAM curves). They are enclosed by a
chaotic region containing two small resonant islands near its top and bottom
edges. In Fig. \ref{octopolar1a} we switch on the octupolar moment
$(\gamma=0.02)$, maintaining the same prolate deformation. The resulting surface
of section presents a more prominent chaotic region, since in this case the
outer zones of resonant islands have overlapped. The  regular regions now
contains ``spindle'' torus. They can be viewed clearly in the central region and
scarcely insinuated in the lateral zone. In Fig. \ref{octopolar1b}, as a
consequence of increase the octupolar moment to $\gamma=0.04$, the chaotic
region is more prominent (the lateral regular zone has disappeared), as well as
the central spindle KAM curves. Finally, Figs. \ref{octupolar2} and
\ref{meridionaloctupolar2} show the effect caused by the progressive rise in the
octupolar deformation, starting from a regular prolate situation.

\begin{figure}
  \epsfig{width=9cm,file=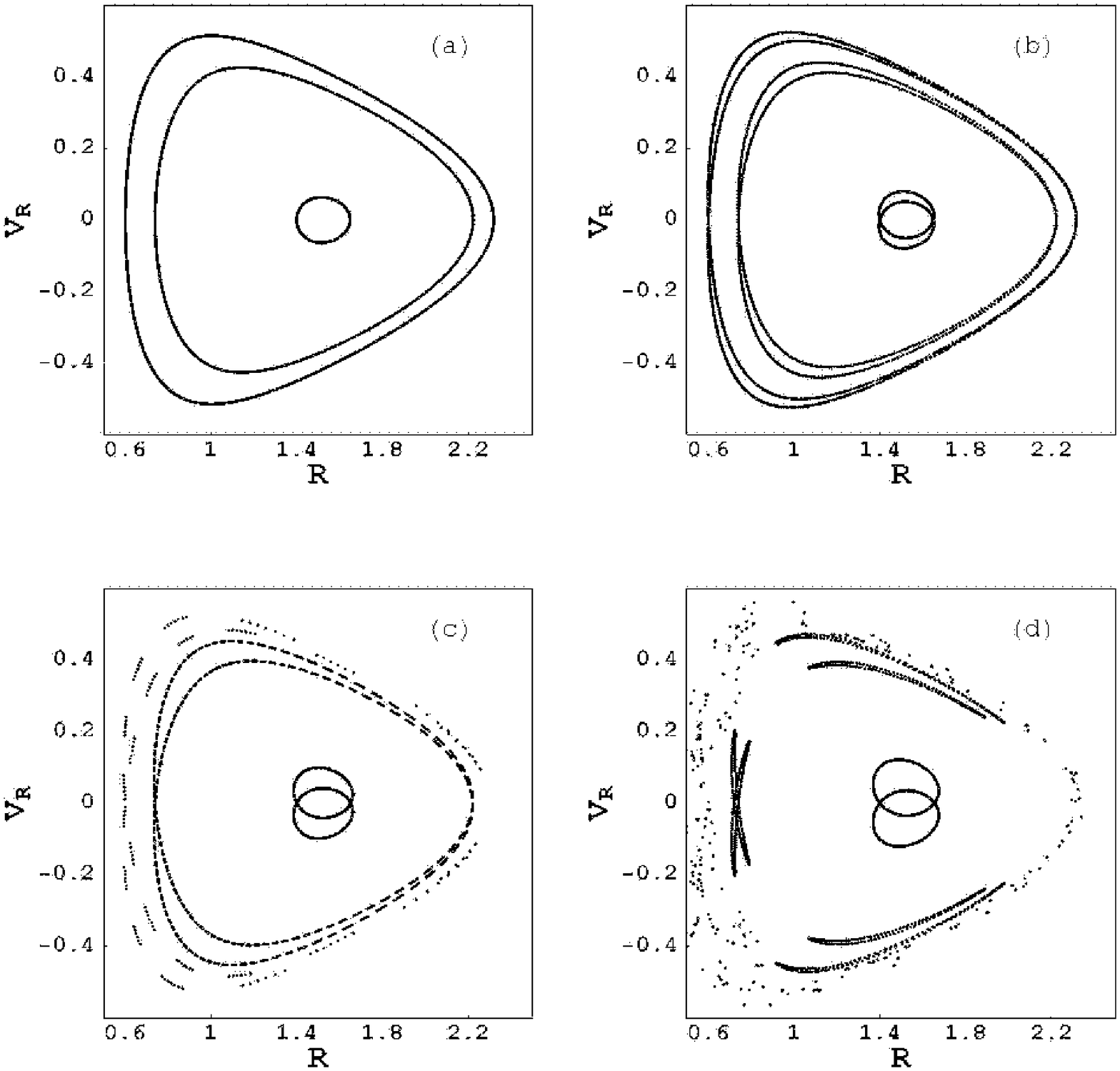}\\
  \caption{Surfaces of section for $\ell=1.1$, $E=-0.32$, $\alpha=1$, $\beta=-0.2$ and
  (a) $\gamma=0$; (b) $\gamma=0.02$; (c) $\gamma=0.04$; (d)
  $\gamma=0.06$. In each case, they are generated by three orbits with initial
  conditions
(i) $z=0$, $R=0.91$, $V_{R}=0$; (ii) $z=0$, $R=0.78$, $V_{R}=0$ and
(iii) $z=0$, $R=1.16$, $V_{R}=0.18$.}\label{octupolar3}
\end{figure}

The fact that by switching on the octupolar moment increases the chaoticity and
leads to apparition of spindle torus, can be viewed even in the case
corresponding to oblate deformation, which commonly presents regular motion (in
fact, there is a wide numerical evidence that particles moving around a monopole
plus an oblate quadrupole are not chaotic). Fig. \ref{octupolar3} shows the
transition from regularity to chaos by increasing $\gamma$. With $E=-0.32$,
$\ell=1.1$ and $\beta=-0.2$, we start from $\gamma=0$ (Fig. \ref{octupolar3}a )
which corresponds to a regular motion. The case $\gamma=0.02$ is also regular
but the KAM curves has been distorted to spindles. In Fig. \ref{octupolar3}c
$(\gamma=0.04)$ the distortion is more prominent and finally, when $\gamma$ has
increased up to $0.06$ we note the apparition of a chaotic region enclosing the
spindles.

\section{Newtonian vs Post-Newtonian Multipolar Moments}\label{1PN}

The post newtonian approximation (commonly called as 1PN approach) gives the
first general relativistic corrections (up to order $v^{2}/c^{2}$, where $c$ is
the speed of light) to the motion equations, when we deal with particles moving
non relativistically ($v\ll c$) in the presence of strong gravitational fields.
We are interested in the situation when a test particle moves around a static
axially symmetric source. As it was shown by Rezania and Sobouti \cite{rezania},
in the stationary  case  we have an integral of motion $E = v^{2}/2 + \Phi +
(2\Phi^{2} + \Psi)/c^{2}$ that can be considered as a generalization of the
classical specific energy. Here, $\Phi$  and $\Psi$ obeys the Poisson equations
\begin{eqnarray}
\nabla^{2}\Phi &=& \frac{4\pi G}{c^{2}}\stackrel{0 \: \: \: \: }{T^{00}},
\label{Poisson1} \\
\nabla^{2}\Psi &=& 4\pi G [\stackrel{2 \: \: \: \: }{T^{00}}+ \stackrel{2 \: \:
\: \: }{T^{ii}}], \label{Poisson2}
\end{eqnarray}
where $\stackrel{0 \: \: \: \: }{T^{00}}$ and $\stackrel{2 \: \: \: \:
}{T^{00}}$ denotes the $0$-th and $2$-th terms in $v/c$ in the $00$-components
of the energy momentum tensor. $\stackrel{2 \: \: \: \: }{T^{ii}}$ represents
the trace of $T^{ij}$ up to order $2$ in $v/c$. Relations (\ref{Poisson1}) and
(\ref{Poisson2}) suggest that both $\Phi$ and $\Psi$ can be expanded in external
multipolar fields. In particular, since $\stackrel{0 \: \: \: \: }{T^{00}}$
represents the source's density \cite{Wein,rezania}, the expansion for $\Phi$ is
the same as in equation (\ref{multipolarNewton}).

In the case of $\Psi$, we can write
\begin{equation}
\Psi = - \frac{\tilde{\alpha}}{\sqrt{R^{2}+z^{2}}} - \frac{\tilde{\beta} (2z^{2}
- R^{2})}{2(R^{2} + z^{2})^{5/2}} - \frac{\tilde{\gamma} (2z^{3} -
3zR^{2})}{2(R^{2} + z^{2})^{7/2}},\label{multipolar1PN}
\end{equation}
where $\tilde{\alpha}$, $\tilde{\beta}$, and $\tilde{\gamma}$ play the role of
the post Newtonian monopole, quadrupole and octupole moments, respectively. In a
similar way as in (\ref{cuadrupoloNew})-(\ref{octupoloNew}), they are given by
\begin{eqnarray}
  \tilde{\alpha} & = & 2\pi\int_{0}^{\infty} r'^{2}dr'\int_{-1}^{1}dx'
  [\stackrel{2 \: \: \: \: }{T^{00}}+
  \stackrel{2 \: \: \: \: }{T^{ii}}],\label{monopoloPosNew}\\
   &  &  \nonumber\\
  \tilde{\beta} & = & 2\pi\int_{0}^{\infty} r'^{4}dr'\int_{-1}^{1}dx'
  P_{2}(x')[\stackrel{2 \: \: \: \: }{T^{00}}+
  \stackrel{2 \: \: \: \: }{T^{ii}}],\label{cuadrupoloPosNew}\\
   &  &  \nonumber\\
  \tilde{\gamma} & = & 2\pi\int_{0}^{\infty} r'^{5}dr'\int_{-1}^{1}dx'
  P_{3}(x')[\stackrel{2 \: \: \: \: }{T^{00}}+
  \stackrel{2 \: \: \: \: }{T^{ii}}],\label{octupoloPosNew}
\end{eqnarray}
where $x'=\cos\theta '=z'/\sqrt{R'^{2}+z'^{2}}$.

The motion equations for a test particle can be derived from the
Hamiltonian
\begin{equation}
H = \frac{1}{2} \left(V_{R}^{2} + V_{z}^{2}\right) + \frac{\ell^{2}}{2R^{2}} +
\Phi + \frac{1}{c^{2}} \left(2\Phi^{2} + \Psi\right), \label{hamiltonian1PN}
\end{equation}
(note that $\ell=R^{2}\dot{\varphi}$ is an integral of motion) and can be
written as
\begin{eqnarray}
\dot{V}_{R} &=& -\frac{\partial U_{ef}}{\partial R}, \label{em3} \\
\dot{V}_{z} &=& - \frac{\partial U_{ef}}{\partial z},\label{em4}
\end{eqnarray}
where we have defined the post Newtonian effective potential
$U_{ef}$ as
\begin{equation}
U_{ef} = \Phi + \frac{1}{c^{2}} \left(2\Phi^{2} + \Psi\right) +
\frac{\ell^{2}}{2R^{2}}.\label{Uef}
\end{equation}
From eqs. (\ref{em3}), (\ref{em4}) and (\ref{Uef}), along with
(\ref{multipolarNewton}) and (\ref{multipolar1PN}), we can study the
trajectories of test particles in a similar way than in the Newtonian case, as
was performed in section \ref{newton}.

\begin{figure}
  \epsfig{width=9cm,file=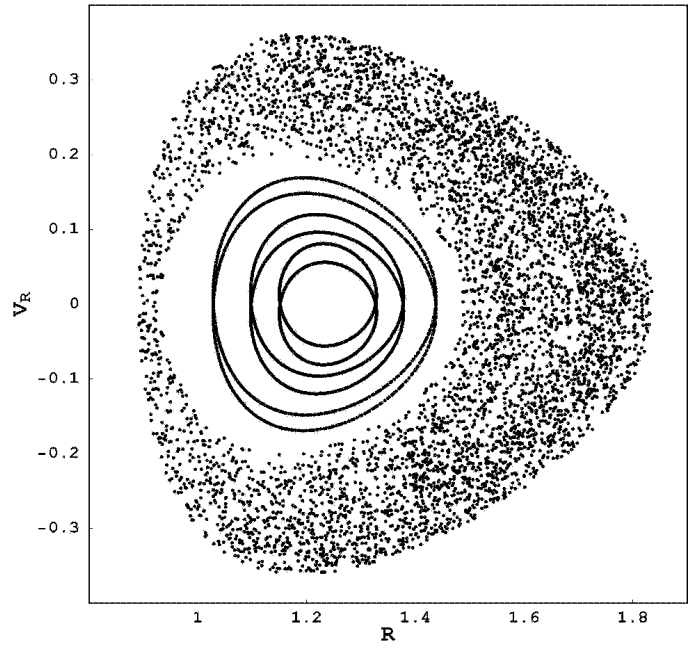}\\
  \caption{In this case, setting the same values of the two previous figures
   but $\gamma=0.04$, we have a prominent chaotic zone enclosing three "spindle" KAM curves.}\label{octopolar1b}
\end{figure}

\begin{figure}
  \epsfig{width=9cm,file=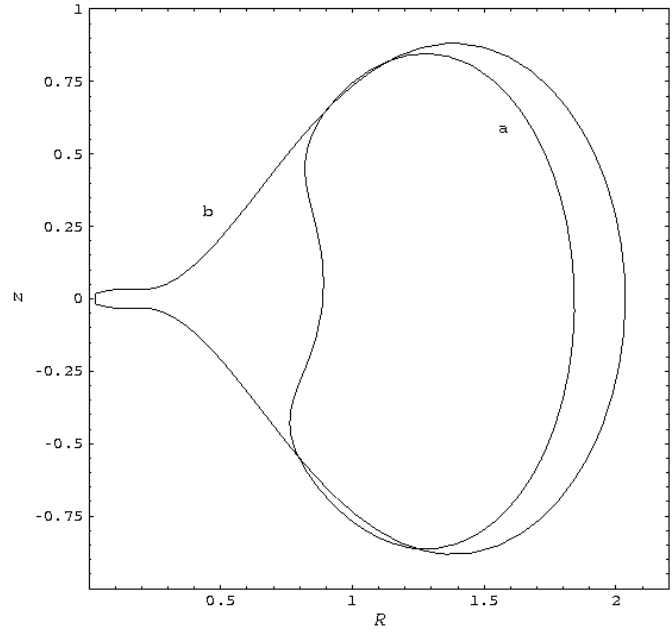}\\
  \caption{Contours of the effective potential $U_{ef}$, for $\alpha=1$, $\beta=0.3$, $\gamma=0.02$, $\tilde{\alpha}=10^{5}$
  and (a) $\tilde{\beta}=-2\times 10^{4}$, $\tilde{\gamma}=10^{4}$; (b)
$\tilde{\beta}=-4\times
  10^{7}$ and $\tilde{\gamma}=-2\times 10^{6}$}.\label{contornos1PNocto}
\end{figure}

\begin{figure}
  \epsfig{width=9cm,file=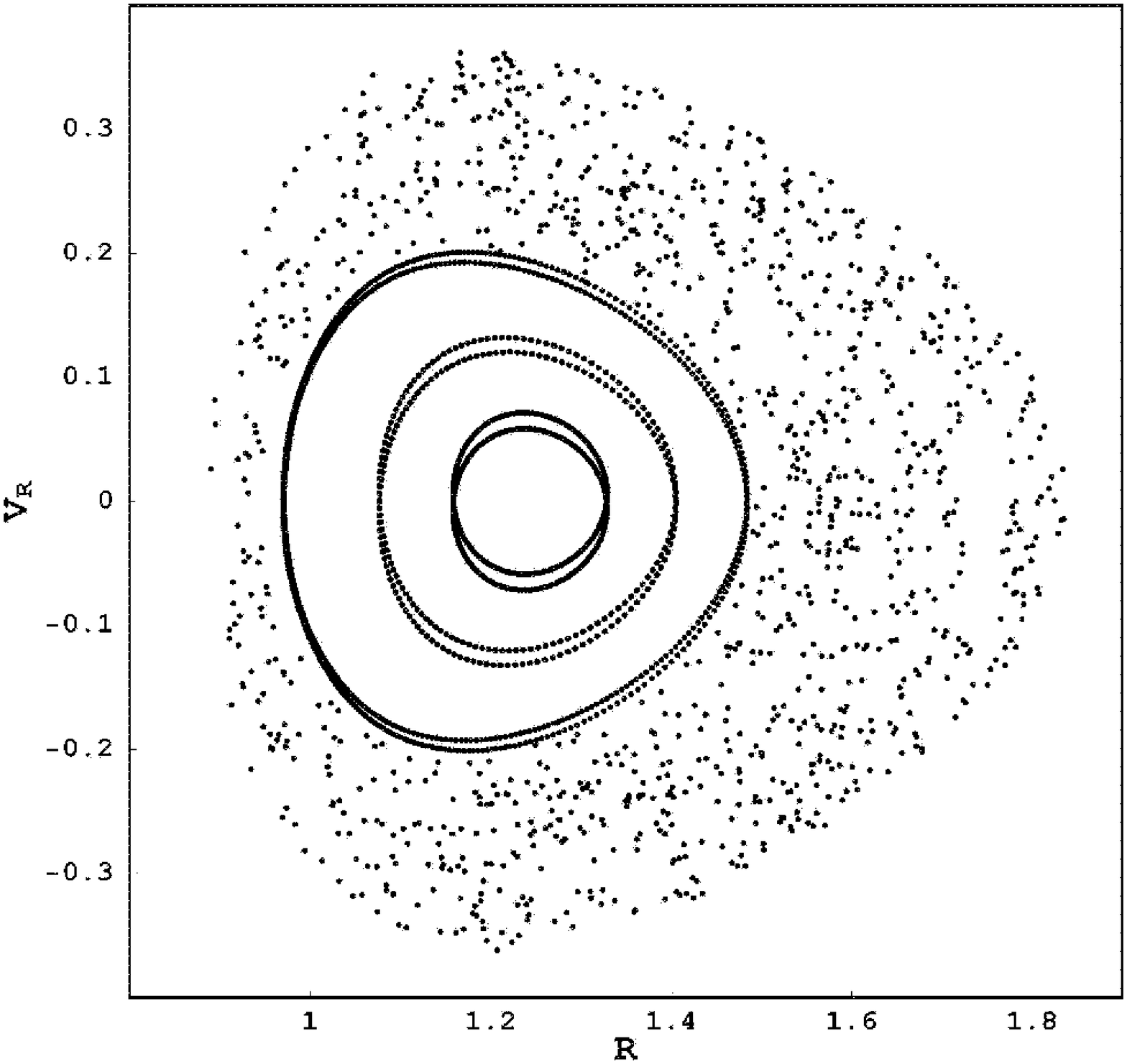}\\
  \caption{In this case, we have chosen the same parameters as in Fig. \ref{octopolar1b} along with
  $\tilde{\alpha}=10^{5}$, $\tilde{\beta}=-2\times 10^{4}$, $\tilde{\gamma}=10^{4}$.}\label{octopolar1PNa}
\end{figure}

\begin{figure}
  \epsfig{width=9cm,file=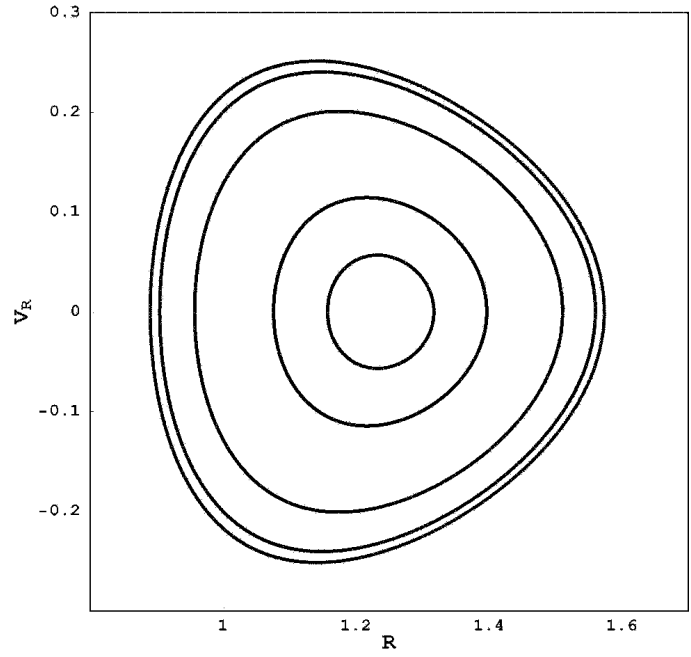}\\
  \caption{Now we keep the same values of $E$, $\ell$, $\alpha$, $\beta$,
  $\gamma$, $\tilde{\alpha}$ as well as the same initial conditions as in previous figure, but $\tilde{\beta}=-4\times
  10^{7}$ and $\tilde{\gamma}=-2\times 10^{6}$. With these values of post Newtonian multipolar moments we have a surface of section
  corresponding to regular motion}\label{octopolar1PNb}
\end{figure}

For example, Fig. \ref{octopolar1PNa} show a $z=0$ surface of section
corresponding to $\ell=0.9$, $E=-0.4$, $\alpha=1$, $\beta=0.3$, $\gamma=0.02$
(the same parameters as in Fig. \ref{octopolar1b}), $\tilde{\alpha}=10^{5}$,
$\tilde{\beta}=-2\times 10^{4}$ and $\tilde{\gamma}=10^{4}$ (according with the
system of units employed, we have to take $c=10^{4}$). It has similar features
as in Fig. \ref{octopolar1b}, but in this case there is not lateral resonant
islands. If one increase adequately the post Newtonian quadrupole and octupole,
an interesting fact occurs: the chaotic zones disappear and the resulting
surface of section reveals completely regular motion. Indeed, as it is shown in
Fig. \ref{octopolar1PNb}, the regular regions are made by rings instead of
spindle torus, for these particular values of multipole parameters. In Figure
\ref{contornos1PNocto} we show the contours of the effective potential
corresponding to these two cases. The former exhibits $z=0$ asymmetry   while
the second is practically symmetric with respect to the equatorial plane.

\section{Concluding Remarks} \label{conclu}

Octupolar deformation in astrophysical objects can introduce significant
modifications to the phase-space structure corresponding to test particles
moving around prolate or oblate centers of attraction. Apart from an increasing
in the chaoticity, the apparition of spindle torus in regular regions is a
remarkable effect caused by the asymmetry of the source with respect to its
equatorial plane. A larger equatorial asymmetry involves  more distorted spindle
KAM curves and more prominent stochastic regions in the phase-space. This fact
carries dramatic consequences in the case of oblate deformed sources, which
usually are associated with regular motion. Here, chaos emerges once we switch
on the octupole moment.

When the first general relativistic corrections are taken into account, through
the post Newtonian approach, the structure in the phase space remains
practically the same as in Newtonian gravity, for typical values of 1PN
multipolar moments. However, there are certain large values for which the test
particle motion becomes regular. This fact introduces an interesting example
where a non-integrable dynamical system becomes integrable through the
introduction of a large perturbation.

\begin{acknowledgments}

J. R-C. and F. L-S want to thank the finantial support from {\it
Vicerrector\'ia Acad\'emica}, Universidad Industrial de Santander.

\end{acknowledgments}

\newpage

\end{document}